\documentclass[pre,twocolumn,showpacs,superscriptaddress]{revtex4}
\newcommand{\figurewidth}{1.\columnwidth}

\usepackage{graphics}
\usepackage{amsmath}
\usepackage{amssymb}
\usepackage{dcolumn}

\begin{document}

\title{Anisotropic Diffusion Limited Aggregation in three dimensions - universality
and non-universality.}

\author{Nicholas R. Goold}
\email{N.R.Goold@warwick.ac.uk}
\affiliation{Department of Physics, University of Warwick, Coventry CV4
7AL, United Kingdom}

\author{Ell\'{a}k Somfai}
\email{ellak@lorentz.leidenuniv.nl}
\affiliation{Department of Physics, University of Warwick, Coventry CV4
7AL, United Kingdom}
\affiliation{Universiteit Leiden, Instituut-Lorentz,
PO Box 9506, 2300 RA Leiden, The Netherlands}

\author{Robin C. Ball}
\email{r.c.ball@warwick.ac.uk}
\affiliation{Department of Physics,
University of Warwick, Coventry CV4
7AL, United Kingdom}

\date{\today{}}

\begin{abstract}
\noindent We explore the macroscopic consequences of lattice anisotropy
for Diffusion Limited Aggregation (DLA) in three dimensions. Simple
cubic and BCC lattice growths are shown to approach universal asymptotic
states in a coherent fashion, and the approach is accelerated by the
use of noise reduction. These states are strikingly anisotropic dendrites
with a rich hierarchy of structure. For growth on an FCC lattice,
our data suggest at least two stable fixed points of anisotropy, one
matching the BCC case. Hexagonal growths, favouring six planar and
two polar directions, appear to approach a line of asymptotic states
with continuously tunable polar anisotropy. The more planar of these
growths visually resemble real snowflake morphologies. 

\noindent Our simulations use a new and dimension-independent implementation
of the Diffusion Limited Aggregation (DLA) model. The algorithm maintains
a hierarchy of sphere-coverings of the growth, supporting efficient
random walks onto the growth by spherical moves. Anisotropy was introduced
by restricting growth to certain preferred directions.
\end{abstract}


\maketitle

\section{Introduction}

The Diffusion Limited Aggregation (DLA) \cite{witten81} model has
been the focus of a great deal of research due both to the fractal
\cite{plischke84, coniglio90, somfai99} and multifractal \cite{amitrano89, jensen03, somfai04} properties of the clusters it produces,
and to its underlying mathematical connection to diverse problems
including solidification \cite{langer80, mullins63}, viscous fingering \cite{nittmann85} and electrodeposition
\cite{matsushita84, brady84}. Its key feature is that the surface irreversibly absorbs an
incident diffusive flux, and growth velocity is locally proportional
to that flux density.

The problem is mathematically ill-posed unless the growth is constrained
to remain smooth below some ``ultra-violet cut-off'' lengthscale, which
in most simulation studies has been supplied by a particle size or
lattice scale. Experimentally the cut-off scale can be more subtle,
for example in solidification regulated by surface tension it varies
with the local incident flux density raised to power $-m$, with $m=1/2$.
Interest has also focussed on the more general Dielectric Breakdown
Model \cite{niemeyer84} cases where growth velocity is proportional
to the incident diffusive flux raised to some power $\eta$. Tuning $\eta$ has been claimed to match appearance between DBM growths (albeit in two dimensions) and real snowflakes \cite{nittmann87}. Recent
theory \cite{ball02prl, ball03pre} suggests quantitative equivalence classes exist in the $\eta$,$m$
plane, so that for example solidification should have a simpler-to-simulate
equivalent at fixed cut-off ($m=0$) at the computational expense
of non-trivial $\eta\ne1$.

A feature of real solidification patterns is that they macroscopically
strongly favour growth in specific directions, corresponding to microscopic
crystal lattice directions. The tendency of snow crystals to grow
six arms is well known, and lately this has been replicated in controlled
laboratory studies \cite{libbrecht99}. Cubic crystalline anisotropy also produces
striking anisotropic ``dendrite'' growth: succinonitryl is the classical
example \cite{huang81, corrigan99}, and lately colloidal crystal exemplars
have been observed in growth under microgravity \cite{russel97, zhu97}.

The manner in which surface tension and its anisotropy select the
morphology of growing tips has been the subject of intense analytical
study \cite{glicksman93}. Full numerical simulations of the continuum growth equations
have confirmed the theory and extended the spatial range out to growths
with significant side-branching \cite{kobayashi93, george02}, but none of these studies could claim
to reach the asymptotic regime of fractal growth.

Simple lattice and particle based simulations differ by having fixed
cut-off scale and lacking realistic local detail, but they can reveal
the limiting behaviour of highly branched growth. In two dimensions
a range of different angular anisotropies have been shown to be relevant
both by theory \cite{meakin87, ball85prl} and through simulations yielding self-similar
dendritic morphologies. Our principle objective in this paper is to
deliver the same level of understanding for three-dimensional simulations,
which have not been systematically explored in the literature to date.

We first introduce a new implementation of the DLA model that involves
enclosing the aggregates with a series of coverings, each made up
of a set of spheres, and show that it can successfully grow large,
three-dimensional DLA clusters. This algorithm entails no intrinsic
lattice or orientational bias, giving us a well posed isotropic reference. 

We then show how anisotropy can be introduced to the algorithm by
confining growth to certain preferred directions, and combine this
with a noise reduction technique in which growth is only permitted
after $H\geq1$ random walkers have hit a growth site on the cluster.
We characterise growths using anisotropy functions which are sensitive
to the growth of fingers along the possible favoured directions. We
present a systematic comparison of the evolution of growths within
cubic symmetry, in particular the respective cases where growth is
favoured along the nearest neighbour directions of one of the simple
cubic, body-centred cubic and face-centred cubic lattices. We also
study growth with uniaxial bias, where growth in polar and planar directions are inequivalently favoured, particularly including the three dimensional hexagonal lattice.

We show that SC and BCC aggregates approach universal, anisotropic
asymptotic states independent of the level of noise reduction, and
that the approach to each of these states follows a single mastercurve.
FCC anisotropy is much slower to emerge, and we show that while high
noise reduction clusters appear to approach an anisotropic fixed point
in the same fashion, the existence of a different fixed point(s) for
low noise reduction growths cannot be ruled out. For growth with uniaxial bias, we observe limiting polar-to-planar aspect ratios of the clusters which depend continuously on the level of input bias.  Thus for the three dimensional hexagonal lattice there appears to be a tunable continuum of asymptotic states.

\section{Growth Algorithm}

The original DLA algorithm takes as its starting point a single fixed
seed particle. A random walker is released from some distance away
and diffuses freely until it hits the seed, at which point it sticks
irreversibly. Further particles are released one at a time and a fractal
cluster is formed. Early simulations were done on (mostly cubic) lattices,
since this reduced the computer run-time required, and cluster sizes
were limited to $N\simeq10^{4}$ particles.

Modern DLA simulations are performed off-lattice and use a number
of tricks to speed up the growth. Since a diffusing particle should
reach the aggregate from a random direction, each walker can be released
from a randomly chosen point on a sphere that just encloses the cluster.
When a walker is far from the cluster it is allowed to take larger
steps than when it is nearby, as long as it never takes a step larger
than the distance to the nearest point of the cluster. A major development
was the Brady-Ball algorithm \cite{ball85jphysa}, which involves covering
the cluster with a series of coarse ``mappings'', to give a
lower bound on the distance to the cluster without looking up the
position of every cluster particle. A further refinement was invented
by Tolman and Meakin \cite{tolman89}, whereby the coarse mappings cover the cluster in
a manner constrained to give a margin of safety: this enables much
simpler (e.g. spherical) moves to be taken. Cluster sizes of $N\simeq10^{7}$
are easily obtainable by these methods. 

Our new algorithm is a fundamentally off-lattice and dimension independent
development of the Brady-Ball-Tolman-Meakin algorithm. We represent the cluster
in terms of a set of \emph{zeroth level} spheres, and we maintain
a hierarchy of coarser scale sphere coverings of these labelled by
higher levels. For simplicity of exposition, we describe the case
where the physical cluster particles are monodisperse, in which case
it is convenient to choose the radius $r_{0}$ of the zeroth level
spheres to correspond to the centre-to-centre distance between contacting
particles (``sticking diameter''). 

Higher level coverings, $n>0$, each consist of a set of spheres of
radius $r_{n}$ such that every zeroth level sphere is \emph{safely}
covered, in the following sense: all points within distance $\phi r_{n}$
of (the surface of) every zeroth level sphere lie inside the covering.
Each covering is also $\textit{simply}$ contained by all higher level
coverings. To make this structure easier to maintain we further required
that each zeroth level sphere was safely covered by a single (not
necessarily unique) sphere at all levels $n>0$. We chose the coverings
to have a geometric progression of size, with $r_{n}=\epsilon^{1-n}r_{1}$,
and terminated the hierarchy when safe covering of the whole cluster
was achieved by a single sphere.

Each sphere at level $n>0$ carries a full set of \emph{downlinks}.
These consist of a pointer to every ``child'' sphere at level
$n-1$ which overlaps the parent. In addition we gave each sphere
(below the highest level) one \emph{uplink}, pointing to one of its
parents; this is only required for the random walks (see later for
choice).

This construction gives an efficient method of generating moves for
our random walkers. At each step we need only determine the highest
level covering that the walker is outside to give a lower bound on
the walker's distance from the cluster. This in turn entails tracking
one (generally not unique) ``enclosing'' sphere which the walker
does lie inside at the next level up.

Given that the walker lies inside an enclosing sphere at level $m$
but outside the lower coverings, we first determine the nearest distance
$d$ from the walker to either the enclosing sphere or any of its
children. The walker can then make a spherically distributed move
of distance $d+\phi r_{m-1}$, because the nearest point of the cluster
must be at least this far away. If the walker has moved outside the
previous enclosing sphere, we follow uplinks until we find a new enclosing
sphere. We then recursively replace that sphere by any of its children
which enclose the walker, until a lowest level enclosing sphere is
found as required for the next move of the walker.

Walkers are deemed to have hit the cluster when they find themselves
inside a cluster particle. They are constrained by a (very small)
minimum step size, typically $10^{-3}r_0$, so they can only ever tresspass
this far into the cluster. They are then ``backed up'' to the cluster
perimeter and added to the aggregate.

As new particles are added to the cluster we must check that they
are safely covered at each level $n>0$. We start at the maximum level,
and create new maximum levels above it if required. Then we move down
levels to $n=1$ checking for safe coverage at each, noting the sphere
that provided this. A level $n$ sphere that safely covers our new
site will necessarily overlap that which did so at the previous level
$n+1$, so the the search at each level can be restricted to the children
of the previous safe container. If none of these give safe coverage
we must add a new sphere at level $n$, ensuring that the integrity
of the data structure is maintained and all the required new links
are put in place.

The safe container at level $n+1$ is made the parent of the new sphere
at level $n$; this is the uplink used by our random walkers. The
new sphere could simply be centred on the particle we wish to add;
however, in an attempt to maximise the efficiency of our coverings
we offset the new sphere by a distance $\gamma r_{n}$ in the direction
of local growth. This offset is constrained by our safe coverage requirement
to obey $r_{0}+\gamma r_{n}+\phi r_{n}<r_{n}$, which is most severe
for $n=1$ leading to \begin{equation}
r_{1}(1-\gamma-\phi)>r_{0}.\label{r1constraint}\end{equation}

We must now find all the spheres which may need a downlink to or from
the new sphere. To facilitate this we impose that each level covering
is simply (but with no required margin) contained within those above
it. In terms of our parameters this requires $r_{n}(1+\gamma)<\phi r_{n+1}+r_{0}$ and
choosing a geometric progression of radii $r_{n}=r_{1}/\epsilon^{n-1}$
with $\epsilon<1$ and no limit on $n$ then requires\begin{equation}
1+\gamma<\frac{\phi}{\epsilon}.\label{constraint}\end{equation}

This constraint ensures that our new sphere is completely covered
by the level $n+2$ safe container, whose child list will hence contain
all the level $n+1$ spheres that need linking to the new sphere.
Similarly, our new sphere is also covered by the level $n+1$ safe
container, and so any level $n-1$ spheres to which the new sphere
needs downlinks are guaranteed to be children $\textit{of the children}$
of that safe container. Thus by remembering the spheres which provided
safe coverage at the previous two levels and selecting parameter values
subject to the constraints (\ref{r1constraint}) and (\ref{constraint})
we can insert all the necessary new links, and ensure the integrity
of our data structure remains intact as the cluster growth proceeds. 

\begin{figure}
\resizebox{\figurewidth}{!}{\includegraphics*{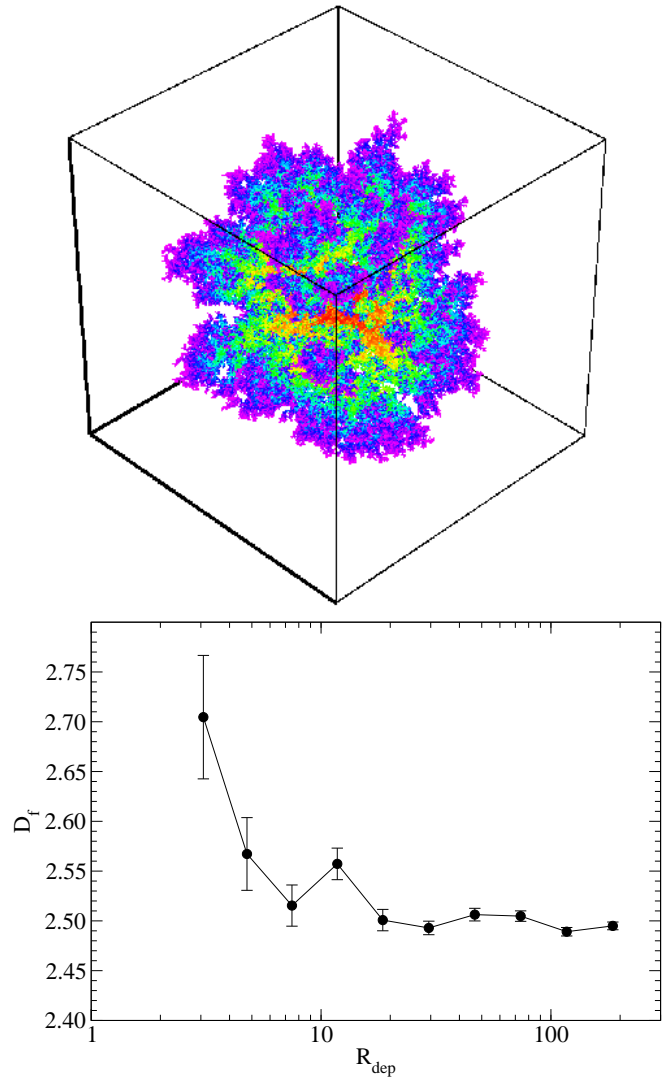}}
\resizebox{\figurewidth}{!}{\includegraphics*{3Ddf.eps}}
\caption{\label{fig:3Dcluster}
(Color online) A three-dimensional DLA cluster grown using the new algorithm containing $N=10^6$ particles, and the fractal dimension $D_f$ plotted against the deposition radius $R_{dep}$ obtained by averaging over a sample of 100 such clusters. $D_f$ converges to a value of $\sim2.5$ in agreement with previous results.
}
\end{figure}

Taking $r_{0}=1$ for convenience, a somewhat ad-hoc optimisation
scheme suggested the following parameters to minimise the run-time
of our program in three dimensions: $r_{1}=2.1$, $\gamma=0.29$, $\phi=0.4$ and $\epsilon=0.3$. We observe that the order of the algorithm is close to linear in $N$, consistent with the earlier discussion of Ball and Brady \cite{ball85jphysa}. Figure \ref{fig:3Dcluster} shows a large off-lattice DLA cluster grown in three dimensions by the new scheme and the convergence of measured fractal dimension
$D_f$ to a value $\sim2.5$ , in good agreement with previous
simulations \cite{tolman89, bowler04}.

\section{anisotropy and noise reduction}

We introduce anisotropy to our simulations by restricting growth to
a set of preferred directions, effectively growing our clusters on
a lattice. When a cluster site is ``grown'' (see following), we add
new ``sticky sites'' of prospective growth offset from the grown site
in each of our lattice directions. Sticky sites in turn grow when
they have accreted a set threshold number of walkers $H\geq1$. Requiring
$H>1$ walkers for growth gives better averaging over the diffusion
field, amounting to a noise reduction. Noise reduction has been widely
used for on-lattice planar DLA simulations \cite{ball02pre}, where it was found
to considerably accelerate the approach to asymptotic morphology.

We have grown aggregates favouring growth in the local simple, body-centred
and face-centred cubic lattice directions. To characterise the macroscopic
anisotropy of a resulting $N$-particle cluster we use functions $A_{K}=\frac{1}{N}\sum_{i=1}^{N}a_{K}(x_{i},y_{i},z_{i})$,
where $(x_{i},y_{i},z_{i})$ are the coordinates of the $i$th particle
relative to the seed of the growth, and $a_{K}$ is a function with
maxima in the appropriate lattice directions. We have constructed
$a_{K}$ out of angular harmonics of order $K$, with the appropriate
symmetry and minimal order to distinguish the different lattice responses
of study.

Growth biassed to the direction of simple cubic axes (relative to the cluster seed) is detected by using a harmonic of
order $4$,\[
a_{4}=\frac{5}{2r^{4}}\left(x^{4}+y^{4}+z^{4}\right)-\frac{3}{2},\]
where $r^{2}=x^{2}+y^{2}+z^{2}$ and the normalisation is chosen such
that $A_{4}=1$ for growth exactly along the lattice axes. Likewise growth along the nearest neighbour directions of an
FCC lattice gives $A_{6}=1$ based on

\begin{eqnarray}
\lefteqn{a_{6}=\frac{112}{13r^{6}}\bigg(x^{6}+y^{6}+z^{6}+} \nonumber\\
& &
\left.\frac{15}{4}\left(x^{4}y^{2}+x^{2}y^{4}+x^{4}z^{2}+x^{2}z^{4}+y^{4}z^{2}+y^{2}z^{4}\right)\right)-\frac{120}{13}. \nonumber
\end{eqnarray}

 The combination of these two enables us to distinguish by sign growth
along SC, BCC or FCC directions as summarised in Table \ref{tab:Avalues},
where values given are for the extreme case of growth confined to
the corresponding nearest neighbour directions from the central seed.

\begin{table}
\caption{ \label{tab:Avalues}
Values of anisotropy functions $A_4$ and $A_6$ for growth along the nearest neighbour directions of simple, body-centred and face-centred cubic lattices.
}
\medskip
\begin{ruledtabular}
\begin{tabular}{cccc}
&  
Simple Cubic &   
Body-centred Cubic & 
Face-centred Cubic \\ \hline
$A_4$ &  
1 &  
-2/3 &  
-1/4  \\
$A_6$ &  
-8/13 &  
-128/117 &  
1  \\
\end{tabular}
\end{ruledtabular}
\end{table}

We have also grown aggregates favouring six planar and two polar directions
of growth. These growths have their polar directions inequivalent
(by any symmetry) to their planar ones so we were naturally led to
admit different values of noise reduction in the two classes of local
growth direction to tune their relative growth. We found the clearest
characterisation of the corresponding growth response of these clusters
simply by measuring their aspect ratios, which we calculate using
extremal radii. We define a cluster's aspect ratio as $z_{\text{max}}/x_{\text{max}}$,
or, in terms of crystallographic notation, $c/a$.

\section{Results}


We grew aggregates favouring SC, BCC and FCC lattice directions at several levels of noise reduction $H$ from $1$ to $100$,
and measured their response using our anisotropy functions $A_{K}$.
The clusters were grown to size $N=3.16\times10^{4}$ particles, where
a site was included in this tally only when it had been hit $H$ times.

\begin{figure}
\resizebox{\figurewidth}{!}{\includegraphics*{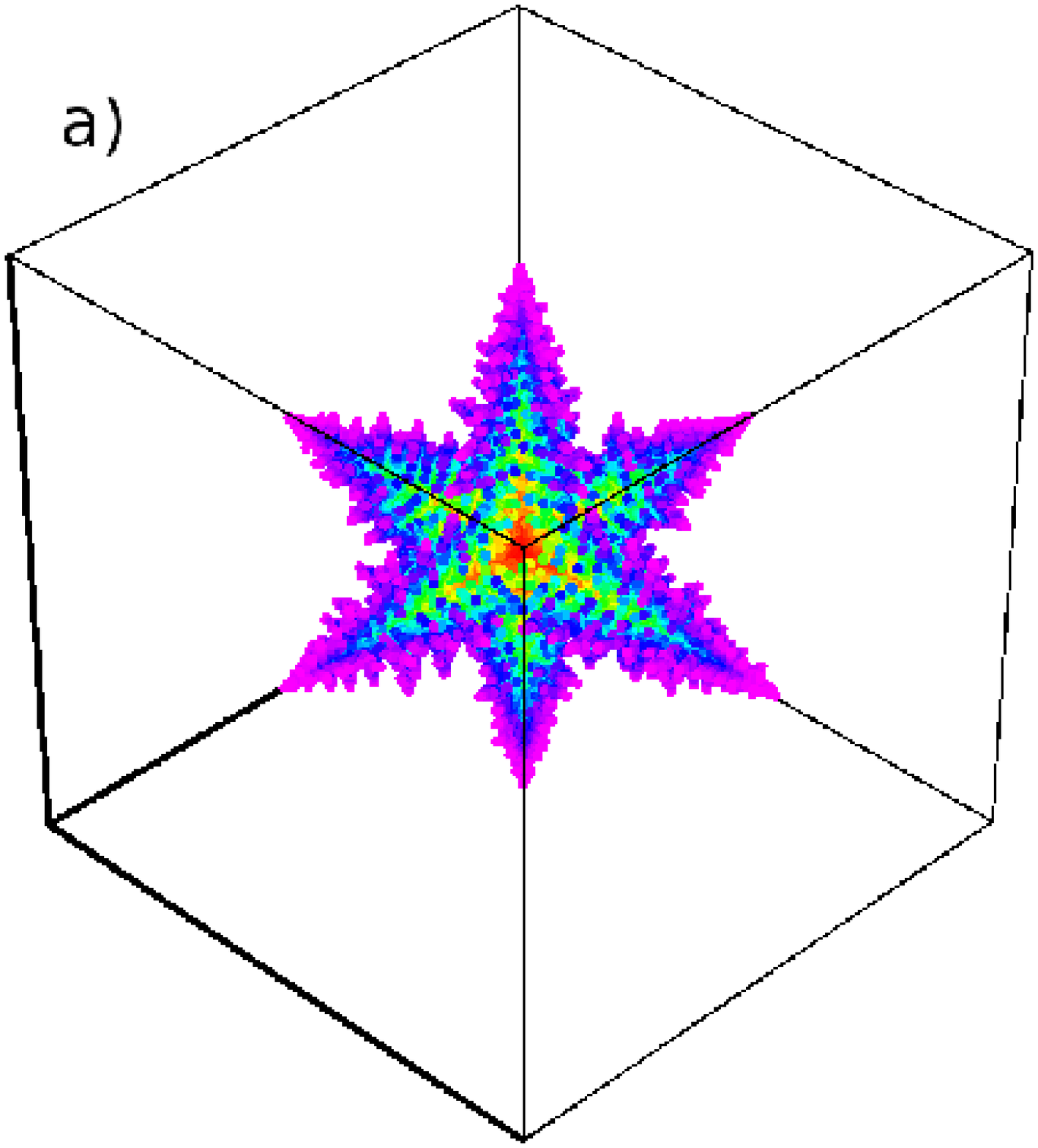}}
\resizebox{\figurewidth}{!}{\includegraphics*{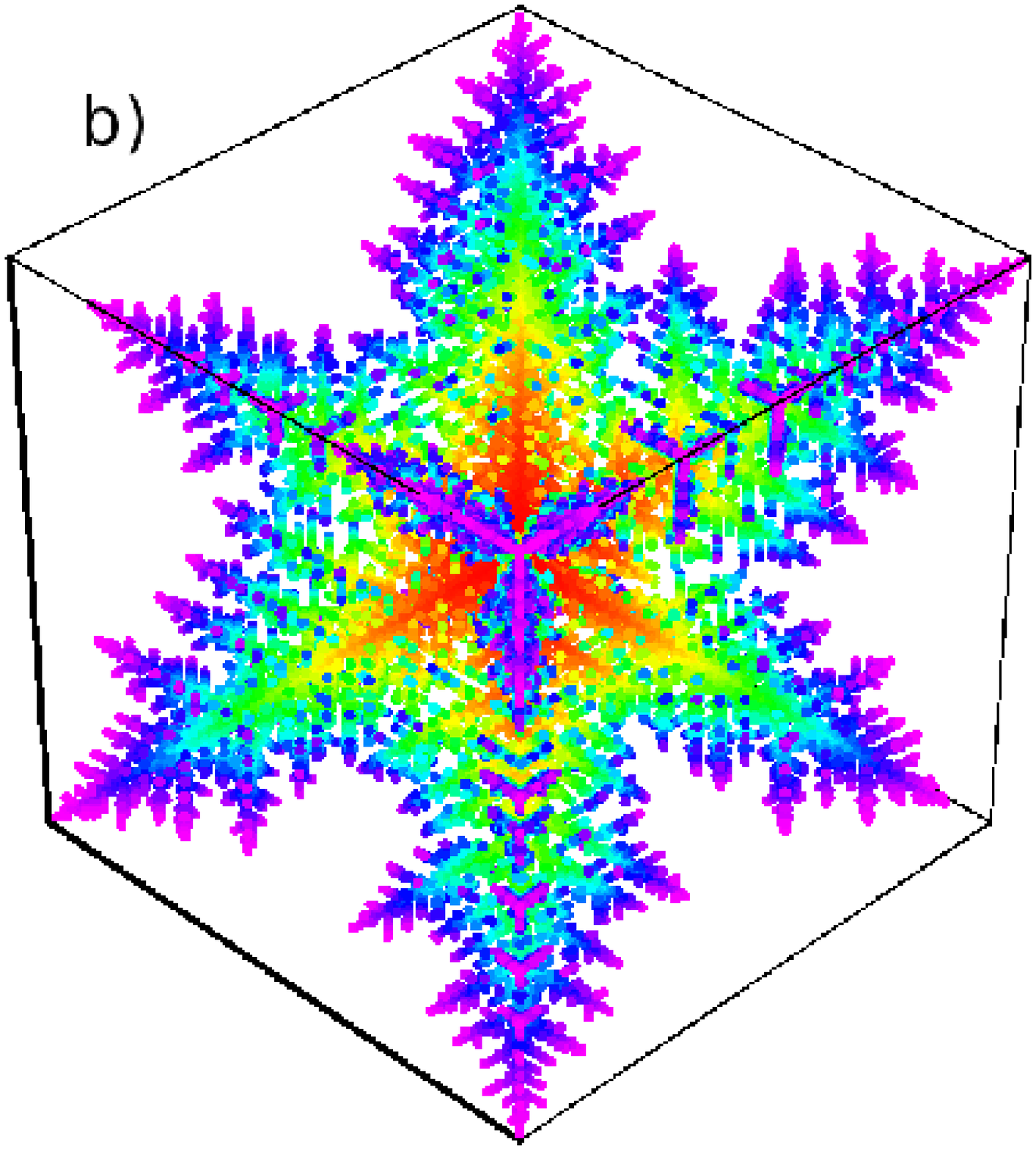}}
\caption{\label{fig:SCBCCpics}
(Color online) Anisotropic DLA clusters grown by the new method: \textbf{a)} simple cubic case, and \textbf{b)} body-centred cubic case. Each contains $3.16\times10^4$ sites, grown under noise reduction such that sites were grown after capturing $H=100$ walkers.
}
\end{figure}

Figure \ref{fig:SCBCCpics} shows example SC and BCC clusters grown at the highest level of noise reduction $H=100$. Both clusters have major arms in the appropriate lattice directions, and each arm exhibits secondary growth along the remaining favoured directions.

\begin{figure}
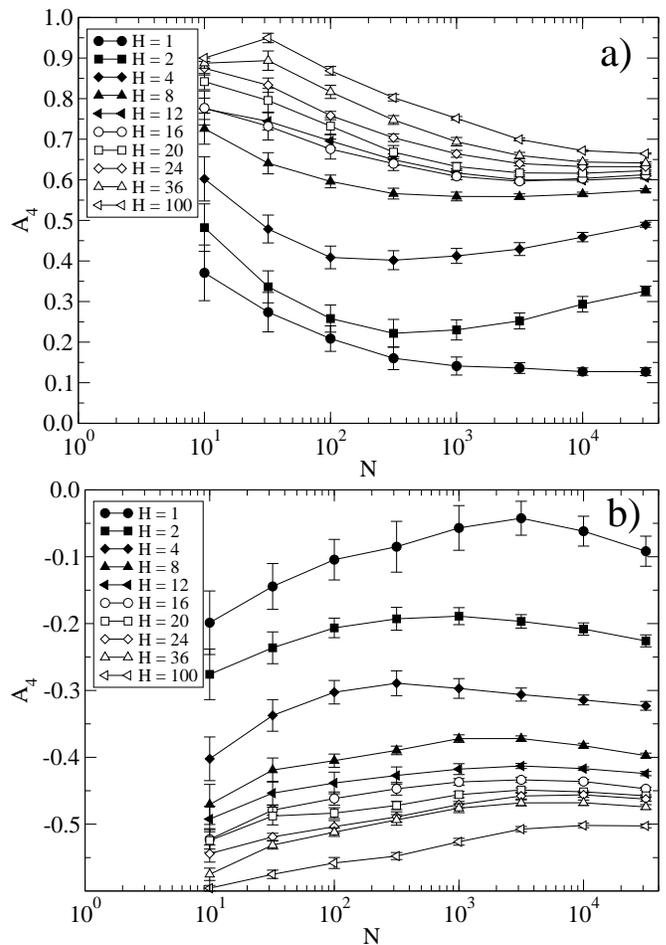

\resizebox{\figurewidth}{!}{\includegraphics*{a4faces.eps}}
\resizebox{\figurewidth}{!}{\includegraphics*{a4corners.eps}}
\caption{\label{fig:A4SCBCC}
Anisotropy function $A_4$ evaluated for \textbf{a)} simple cubic and \textbf{b)} body-centred cubic growths at various levels of noise reduction $H$ as a function of the number of sites grown $N$.  Each curve is based on the average over 10 clusters.  Comparing to reference values in Table \ref{tab:Avalues}, these confirm quantitatively the visual impression from Figure \ref{fig:SCBCCpics} that the respective SC and BCC anisotropies are self-sustaining under growth.
}
\end{figure}

The measured anisotropy $A_4$ for SC and BCC growths at various $H$ is shown in Figure \ref{fig:A4SCBCC}.
Both sets of clusters appear to approach universal asymptotic values
of $A_{4}$ independent of noise reduction: $A_{4}(\infty)\simeq0.65$
for SC growths and $A_{4}(\infty)\simeq-0.5$ for BCC growths. These
values can be approached from both above and below, depending on $H$.

\begin{figure}
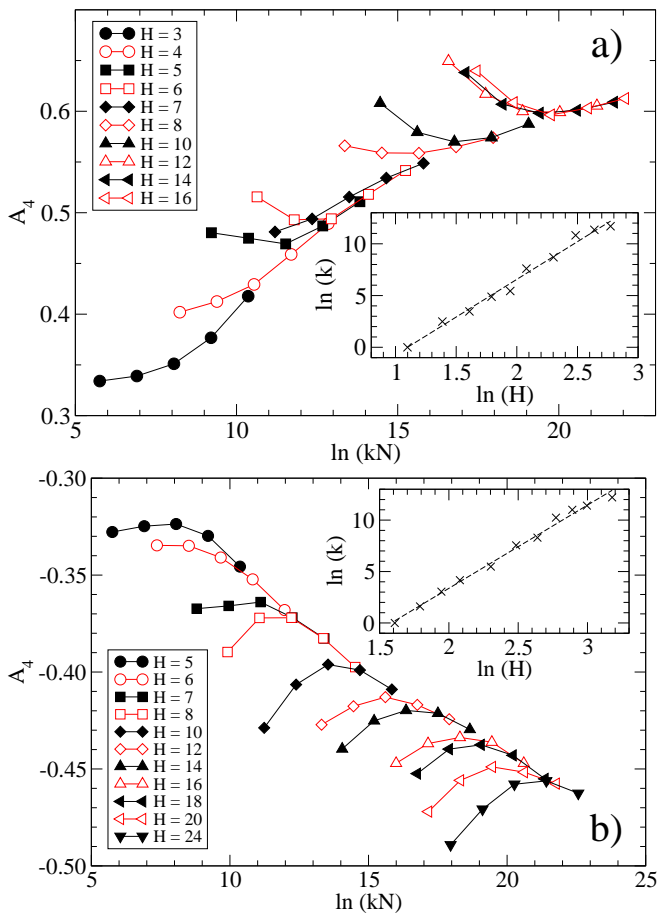

\resizebox{\figurewidth}{!}{\includegraphics*{shifta4faces.eps}}
\resizebox{\figurewidth}{!}{\includegraphics*{shifta4corners.eps}}
\caption{\label{fig:masterSCBCC}
(Color online) Mastercurves of the evolution of $A_4$ for \textbf{a)} simple cubic and \textbf{b)} body-centred cubic growths. These suggest universal approach to respective SC and BCC fixed points.  The insets show how the shift factors $k$ applied to $N$ vary with noise reduction parameter $H$.
}
\end{figure}

The consistency of the shapes of the anisotropy curves suggest that
for these types of growth there may exist ``mastercurves'' that embody
the evolution of $A_{4}$ towards its asymptotic value, as a function of rescaled $N$. All individual
curves, regardless of $H$, will lie somewhere on these mastercurves.
To test this hypothesis for each case we shifted the curves along
the $N$-axis by a factor $k(H)$ until, by eye, they appeared to
follow a single curve. Figure \ref{fig:masterSCBCC}(a) and (b)
shows the results of this procedure for both the SC and BCC growths.
For each case, we have used only the results for $N>10^{2}$ in order
to be sure of the correct general trend, and we could not use the very
low $H$ curves because they vary too little across the simulation
range to give sufficient vertical overlap.
The figure shows power law relationships between the noise reduction
$H$ and the shift factors $k(H)$ in both cases, further evidence
that this mastercurve approach correctly describes the evolution of
SC and BCC growths. 

In the SC case, Figure \ref{fig:masterSCBCC}(a), the shifted curves for values of
$H$ from $3$ to $16$ are shown. For $H>16$, since the anisotropy
curves are very close to the asymptotic value of $A_{4}$ and are hence
very flat, this curve-shifting process fails. There presumably exists
some ideal noise reduction value $H^*$ for which the $A_{4}$ curve
will approach the asymptote most quickly, and we would of course expect
the power law scaling to break down as $H$ approaches this value.
The curves for very high values $H\geq28$ approach the asymptotic
value from above, and it should presumably be possible to map them
onto a second mastercurve. However to test that systematically would require more data of considerable computational cost.

For the BCC case, Figure \ref{fig:masterSCBCC}(b), the anisotropy is much slower
to emerge from the noise and the curves for all values of $H$ save
the very highest $H=100$ approach the asymptotic value from the same
direction, and the mastercurve includes all values of $H$ from 5
to 24. Above $H=24$, the procedure fails in the same fashion as the
SC case as $H^*$ is approached. 

\begin{figure}
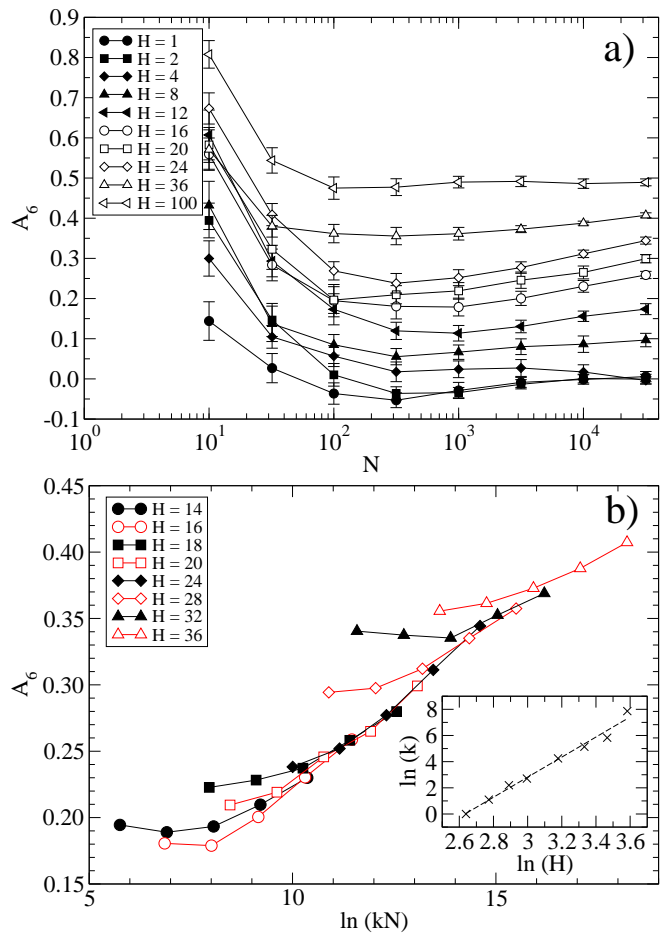

\resizebox{\figurewidth}{!}{\includegraphics*{a6edges.eps}}
\resizebox{\figurewidth}{!}{\includegraphics*{shifta6edges.eps}}
\caption{\label{fig:anisFCC}
(Color online) \textbf{a)} Anisotropy function $A_6$ evaluated for FCC growths, based on the average over 10 clusters per curve. Only at higher noise reduction levels is there clear indication that the FCC anisotropy is sustained under growth.
\textbf{b)} Mastercurve for FCC growths at $H\ge 14$, which do appear to exhibit a common evolution, with the corresponding shift factors inset.
}
\end{figure}

Anisotropy curves $A_{6}$ for FCC growths are shown in Figure \ref{fig:anisFCC}(a),
and it is immediately apparent that their behaviour is not as straightforward
as the SC and BCC cases. For high $H$ growths, $A_6$ appears to be increasing
in a fashion similar to that previously observed, suggesting the existence
of a fixed point of anisotropy for FCC growth at $A_{6}(\infty)\simeq0.48$.
All the curves approach this value from below, suggesting that the
FCC anisotropy is much slower to emerge from the noise than the SC
and BCC anisotropies. This seems reasonable given that
the FCC anisotropy has more competing ``arms'' than the other growths,
and we have verified that the $H=100$ clusters do indeed appear to
have a full set of 12 arms. A mastercurve for these higher values
of $H$ is shown in Figure \ref{fig:anisFCC}(b), and seems to describe
these results well. 

\begin{figure}
\resizebox{\figurewidth}{!}{\includegraphics*{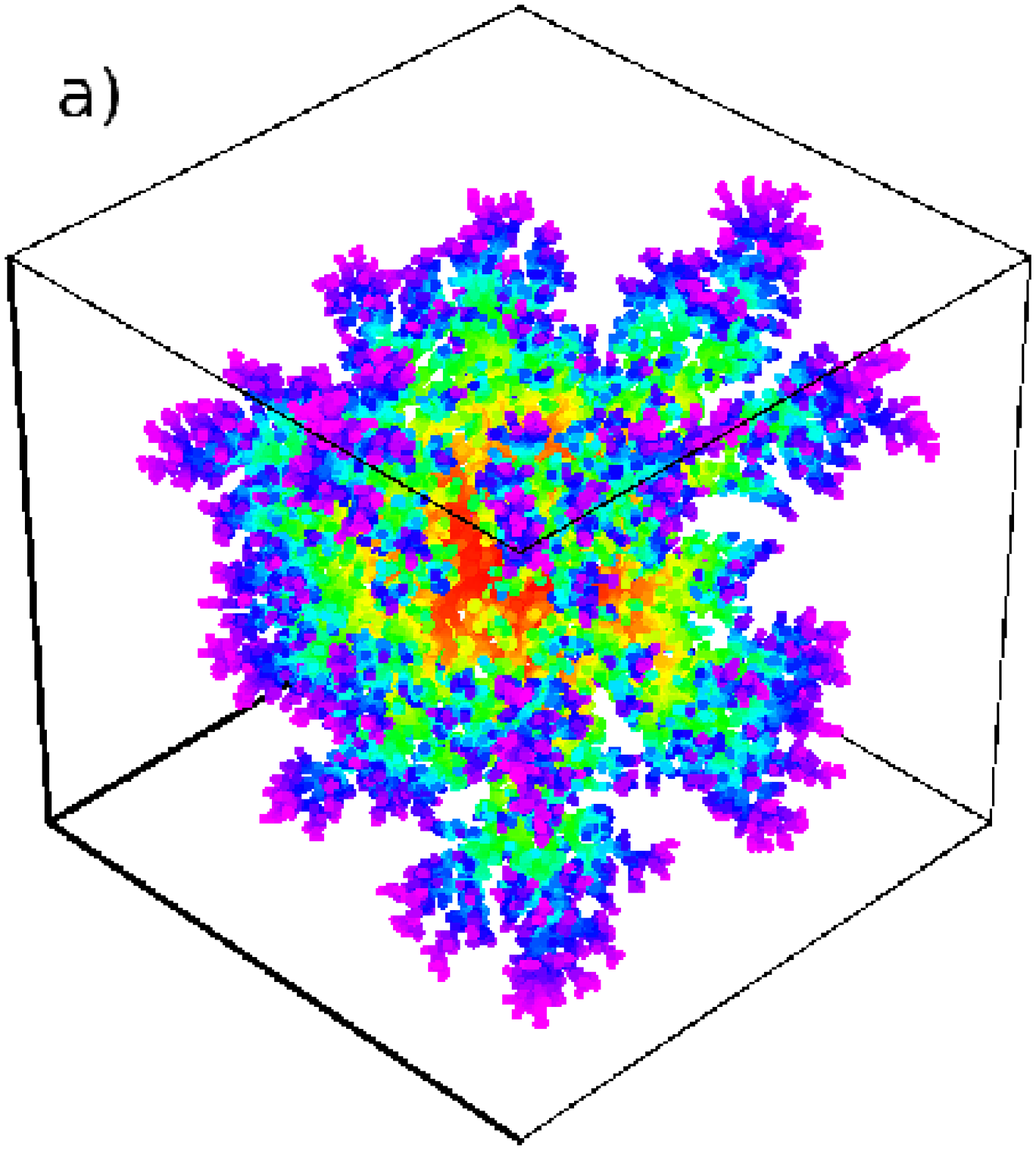}}
\resizebox{\figurewidth}{!}{\includegraphics*{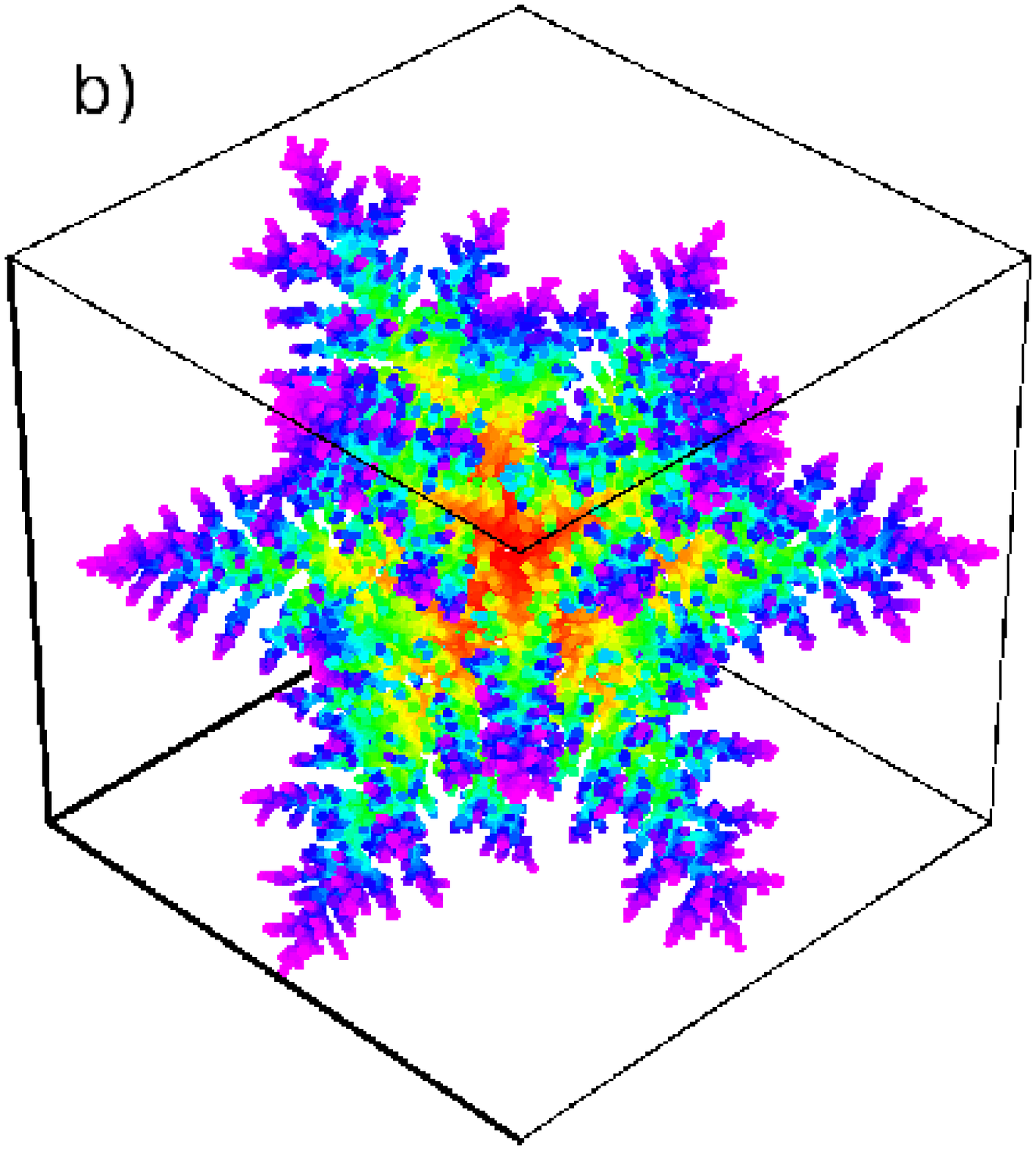}}
\caption{\label{fig:FCCpics}
(Color online) FCC clusters grown by the new method: \textbf{a)} low noise reduction $H=6$, and \textbf{b)} high noise reduction $H=100$. Both clusters contain $3.16\times10^4$ sites grown. The low noise reduction case appears to show some growth bias to the BCC lattice directions (corners) as per BCC lattice growth in Figure \ref{fig:SCBCCpics}. The high noise reduction cluster exhibits growth in all twelve FCC lattice directions, which correspond to the mid-edges of the box drawn.
}
\end{figure}

For low noise reduction clusters $H<8$ however, $A_{6}$ does not
increase over the course of the growth, and if anything appears to
be \emph{decreasing} at large $N$ towards a value of about zero,
suggesting the possible existence of another fixed point. Visualisations
of these low $H$ FCC clusters appeared to indicate some growth along
the BCC lattice directions; Figure \ref{fig:FCCpics} shows an example of this for a low noise reduction $H=6$ cluster, and for comparison a high noise reduction $H=100$ cluster exhibiting some growth in all 12 FCC lattice directions. 

\begin{figure}
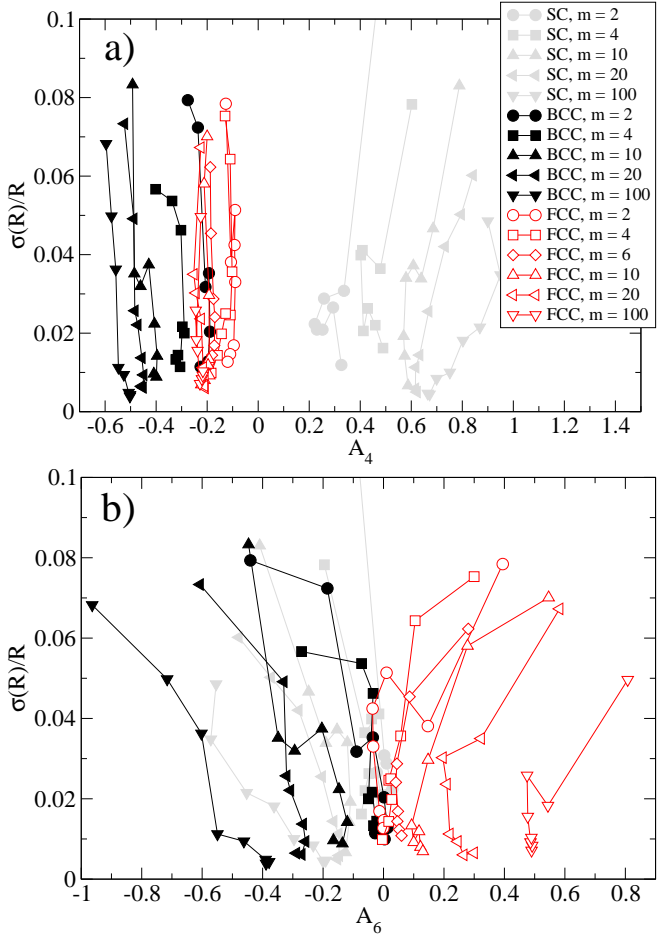

\resizebox{\figurewidth}{!}{\includegraphics*{fluca4.eps}}
\resizebox{\figurewidth}{!}{\includegraphics*{fluca6.eps}}
\caption{\label{fig:fluc}
(Color online) The relative fluctuation in cluster radius (at fixed $N$), plotted against each of the anisotropy measures $A_4$ and $A_6$ as clusters grow.  Increasing $N$ corresponds to moving generally downwards in these plots, and the symbols are the same on both panels.  
}
\end{figure}

We were hence led to apply the BCC anisotropy
function $A_{4}$ to the FCC aggregates, and for comparison $A_{6}$
was also evaluated for the BCC and SC clusters. Studies of two-dimensional
anisotropic DLA \cite{barker90} have had some success focussing
on the interplay between anisotropy and noise in the growth process, and in this spirit we measured $\sigma(R)/R$, where $R$ is the
deposition radius of a cluster and $\sigma(R)$ is the standard deviation
of this measurement. This quantity offers a simple measure of
fluctuations due to noise during cluster growth; $\sigma(R)/R$
is plotted against $A_{4}$ in Figure \ref{fig:fluc}(a) and against $A_{6}$ in
Figure \ref{fig:fluc}(b) for clusters of each type at various $H$.

Figure \ref{fig:fluc}(a) shows that for all clusters the noise decreases
reasonably monotonically as growth proceeds. SC and BCC growths for
all $H$ can be seen to converge towards their respective fixed point
values of approximately $0.65$ (SC) and $-0.5$ (BCC). The FCC growths
are all grouped around $A_{4}\simeq-0.2$, and this plot fails to
explain the behaviour of the low $H$ FCC clusters. However, Figure
\ref{fig:fluc}(b) gives us an idea of what may be happening: whereas
the higher $H$ growths head towards the same final value ($A_{6}\simeq0.5$),
the curves for the low $H$ FCC clusters have not turned towards this
value and appear to be following similar trajectories to the low $H$
BCC clusters. All the SC and BCC clusters appear to be approaching
common asymptotic values of $A_{6}$ of approximately $-0.2$ and $-0.35$
respectively.

\begin{figure}
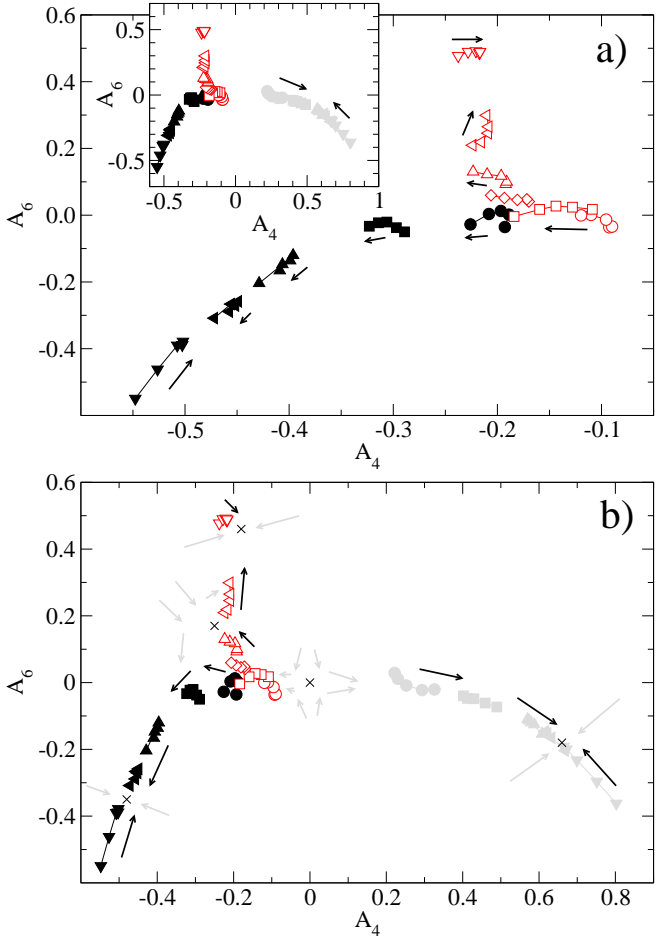

\resizebox{\figurewidth}{!}{\includegraphics*{acomp.eps}}
\resizebox{\figurewidth}{!}{\includegraphics*{schematic.eps}}
\caption{\label{fig:acomp}
(Color online) \textbf{a)} The evolution of clusters in the plane of $A_4$ and $A_6$ for BCC and FCC clusters at various noise reduction levels (always averaging over samples of 10), with arrows to indicate the direction of trajectories. The inset shows the well separated evolution of SC clusters in the same plane.
\textbf{b)} The same data can be interpreted as renormalisation flows for how effective parameters evolve as a function of lengthscale, leading to the inferred fixed points shown (crosses). It is an assumption here that the lowest relevant angular harmonics $A_4$, $A_6$ do capture the key distinction between the three different applied anisotropies. Bold arrows show observed evolution whereas gray arrows show the presumed flow from other starting points. The symbols on both panels are the same as those in Figure \ref{fig:fluc}.
}
\end{figure}

Further evidence for this explanation of the FCC cluster behaviour
is given by plotting $A_{4}$ against $A_{6}$ for BCC and FCC growths in
Figure \ref{fig:acomp}(a). This clearly shows the BCC clusters evolving
(in a direction dependent on $H$) towards a fixed point. The high
$H$ FCC growths also head to their own fixed point, whereas the low
$H$ growths are moving in a different direction, towards the BCC
fixed point. The inset of Figure \ref{fig:acomp}(a) shows the position
of the SC growths in the $A_{4},A_{6}$ plane; they can also be seen
to approach a fixed point from different directions depending on $H$. 

This information allows us to build what we believe to be a consistent
picture of the evolution of all three types of growth, interpreted in terms of how the parameters $A_4$ and $A_6$ evolve as a function of increasing lengthscale. This is shown in Figure \ref{fig:acomp}(b). Our anisotropy curves have shown the existence of stable fixed points, for each of SC, BCC and high noise reduction FCC clusters: assuming that the variables $A_4$ and $A_6$ capture the key distinction between the different anisotropies studied, these directly imply the three separate stable fixed points shown on Figure \ref{fig:acomp}(b). There is presumably an unstable fixed point located at $(0,0)$ in
the $A_{4},A_{6}$ plane corresponding to isotropic growth, and the
differing trajectories of the FCC clusters dependent on $H$ implies
the existence of another unstable fixed point to separate the two
behaviours. The measured directions of our data allow us to predict trajectories of growths with different starting points, with flow away from or towards each fixed point depending on its nature.

\begin{figure*}
\resizebox{\figurewidth}{!}{\includegraphics*{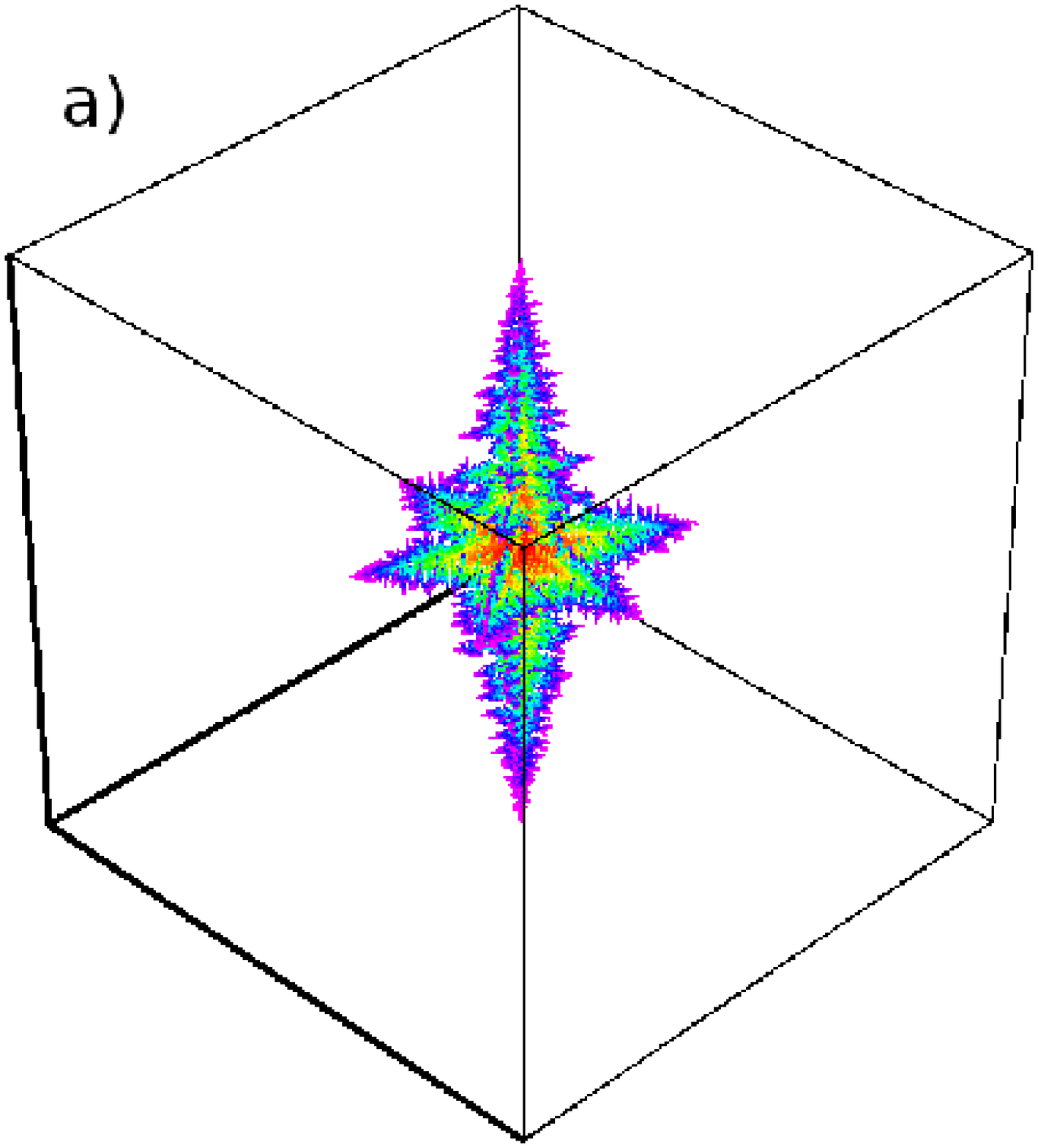}}
\resizebox{\figurewidth}{!}{\includegraphics*{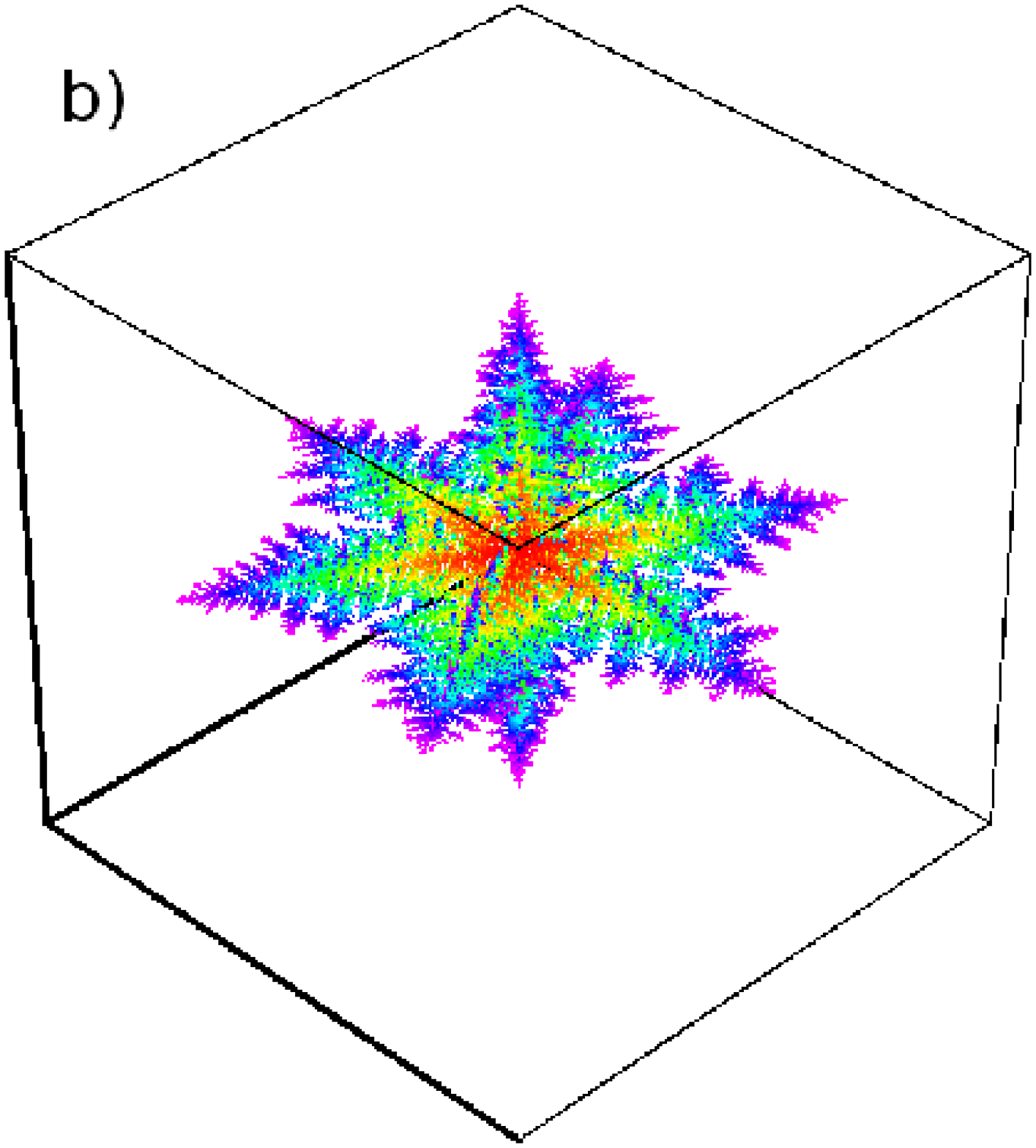}}
\resizebox{\figurewidth}{!}{\includegraphics*{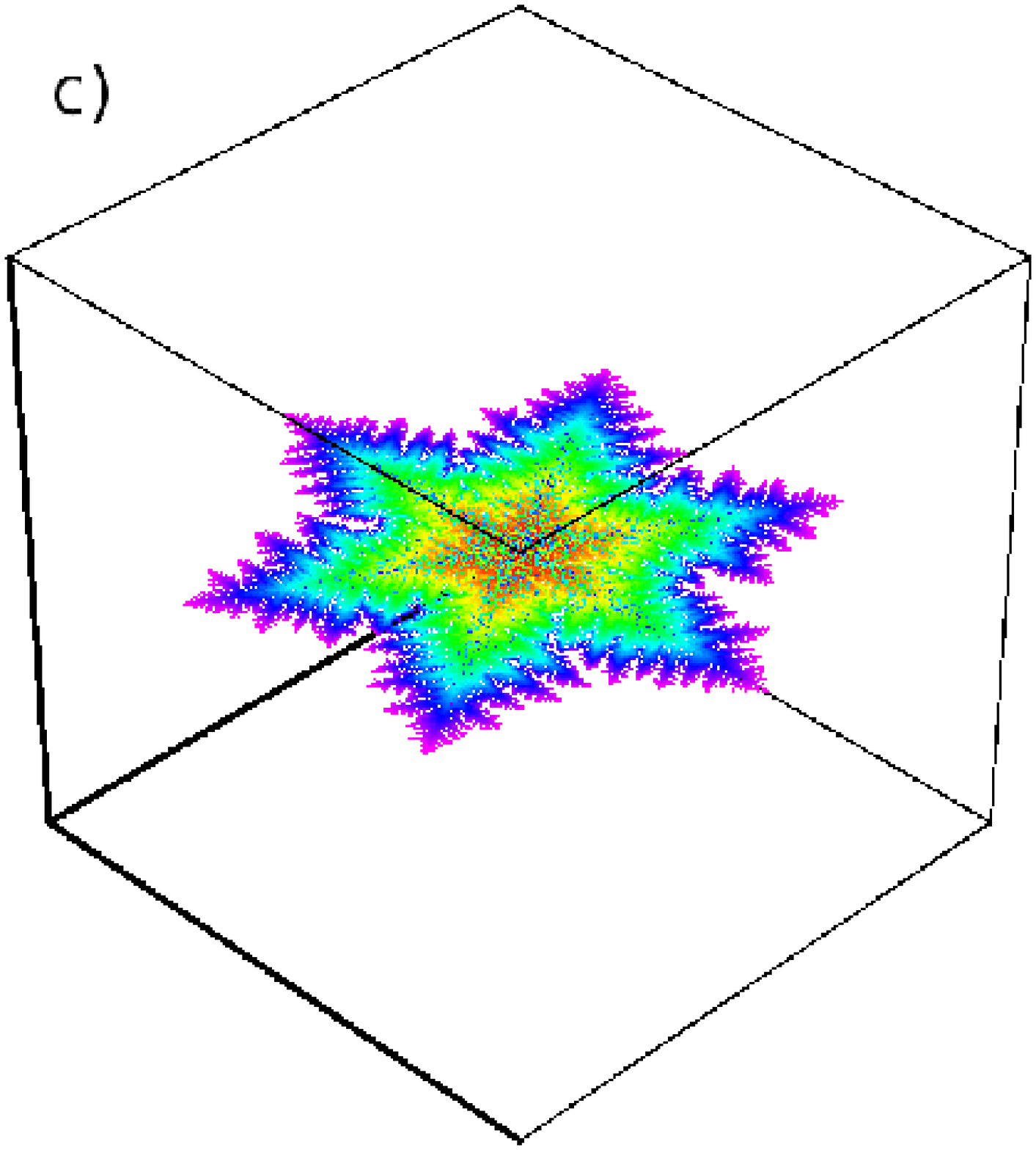}}
\resizebox{\figurewidth}{!}{\includegraphics*{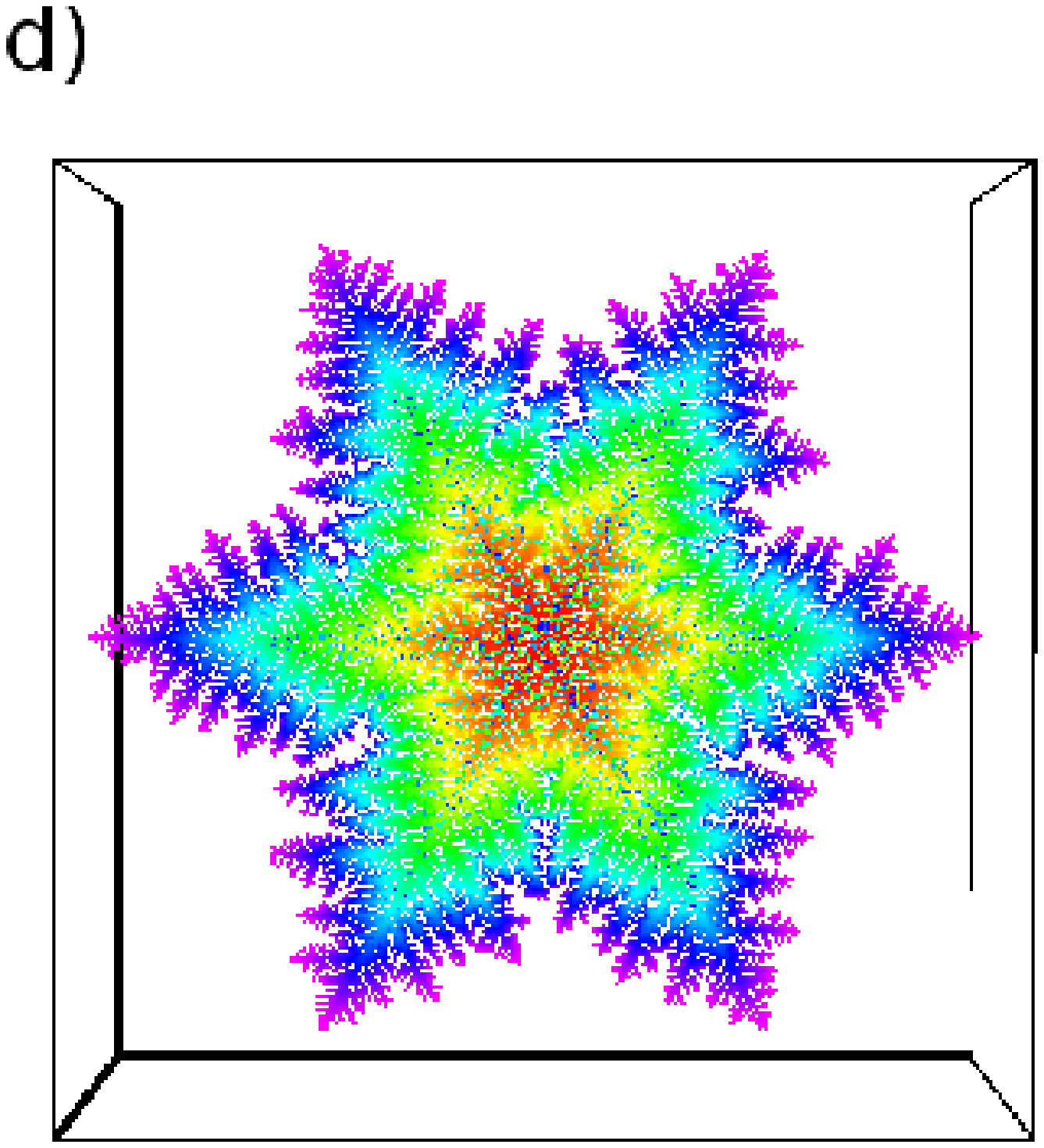}}
\caption{\label{fig:hexpics}
(Color online) Clusters grown favouring six planar directions and two polar directions. The parameter $p$ is a measure of the relative ease of planar growth compared with polar, with high values of $p$ favouring planar growth. Shown are \textbf{a)} $p=1$, \textbf{b)} $p=1.5$, and \textbf{c)} $p=4$; \textbf{d)} shows the $p=4$ cluster viewed from above, highlighting the complex six-armed morphology of these growths.
}
\end{figure*}

We next turn our attention to hexagonal growths. We define a parameter
$p=\frac{H_{z}}{H_{xy}}$ where $H_{\text{z}}$ is the
number of walker hits required to grow a site in either of the favoured
polar directions and $H_{\text{xy}}$ is the number of hits required
for growth in a favoured planar direction. For all growths $H_{\text{xy}}=100$
while $H_{\text{z}}$ was varied between simulations to give values
of $p$ ranging from $0.5$ to $4$. Low values of $p$ produced column-like
growths, while high $p$ resulted in virtually flat aggregates with
six arms in the plane. Example clusters are shown in Figure \ref{fig:hexpics}.


\begin{figure}
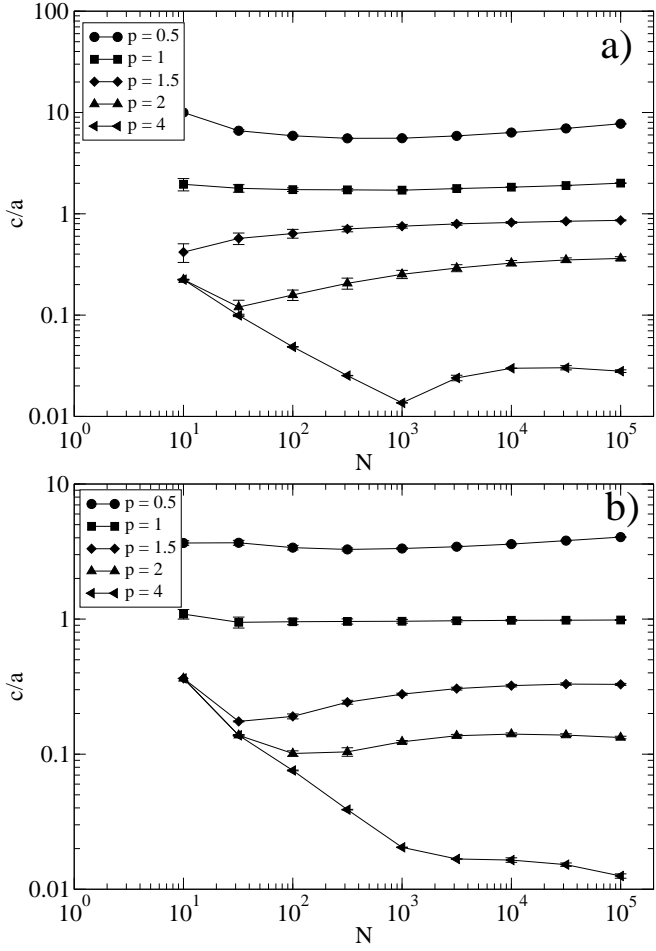

\resizebox{\figurewidth}{!}{\includegraphics*{avaspratnr100hex.eps}}
\resizebox{\figurewidth}{!}{\includegraphics*{avaspratnr100faces.eps}}
\caption{\label{fig:asp}
Aspect ratios of \textbf{a)} hexagonal growths, and \textbf{b)} SC growths, both with tunable input polar anisotropy $p$. The data is from 5 clusters of each type for each value of $p$; using both polar arms and the 6 (hexagonal) or 4 (SC) planar arms means each cluster provides 12 or 8 measurements, respectively, of an aspect ratio at size $N$.
}
\end{figure}

Aspect ratios of these hexagonal clusters are shown in
Figure \ref{fig:asp}(a). The results are from 5 clusters of size $N=10^{5}$
at each value of $p$, although since a cluster possesses six planar and two polar arms each provides us with twelve measurements of an aspect ratio. Strikingly, the aspect ratios remain almost
constant for $N\geq10^{3}$, suggesting the existence of a continuous
spectrum of fixed points which depend on the input anisotropy. This
is contrary to a simple expectation of two fixed points, favouring
polar or planar growth respectively.

To investigate this interesting result further we tuned the polar
growth of some SC clusters in the same way. Since these clusters possess
fewer competing arms than the hexagonal growths, for the same size
$N$ they should be more converged towards their asymptotic states. The aspect
ratios of these clusters are shown in Figure \ref{fig:asp}(b), and they
appear very similar to the hexagonal results. They do suggest a coherent
explanation, however. The $p=1$ clusters are of course just the standard
SC clusters investigated above and since polar and planer lattice
directions are in this case equivalent, they unsurprisingly display
a constant aspect ratio of 1. This fixed point is presumably stable under growth.
The aspect ratios of the extreme cases $p=0.5$ and $p=4$ appear
to diverge at very large $N$, suggesting the existence of additional
fixed points at infinity and zero, corresponding to column-like or
flat growths respectively. The intermediate growths $p=1.5$ and $p=2$
approach constant aspect ratios which we interpret to be the fixed
point at 1 displaced by the input tuning of the growths. The striking
similarity between these results and those for the hexagonal clusters
leads to a similar explanation for their origin.

\section{Conclusions}

We have utilised an efficient, dimension-independent numerical implementation
of the DLA model to explore the effects of several different lattice
anisotropies on three-dimensional aggregates.

For cubic anisotropies we used functions with maxima in the appropriate
lattice directions to characterise our aggregates. We have shown that
SC and BCC growths approach universal asymptotic states, independent
of the level of noise reduction, and that in each case the evolution
of the cluster anisotropy can be described by mastercurves.

For face-centred cubic anisotropy, high noise reduction clusters also
appear to approach a common asymptotic state. By evaluating the clusters'
anisotropic response in the BCC lattice directions we have shown that
lower noise reduction FCC growths appear to evolve towards the BCC fixed
point. The final appearance of these low noise reduction FCC clusters remains
uncertain.

We also studied hexagonal anisotropies with six favoured planar directions
and two favoured polar directions. We tuned our growths by varying
the criterion for growth in a polar direction, and somewhat surprisingly
found that the aspect ratios of our clusters appear to exhibit a continuous
spectrum of final states, dependent on the input tuning. These growths
bear a striking resemblence to snowcrystal morphologies.

Accurate simulation of real solidification patterns like snowcrystals
involves including the effect of a non-constant small-scale cutoff.
Theory \cite{ball02prl, ball03pre} suggests an equivalence between this and simple dielectric
breakdown model growth. A new method of realising DBM growth using
random walkers has been implemented in two dimensions \cite{somfai04},
and our new code is ideally suited to extending this investigation
to three dimensions, of which little or nothing is known. Combining
these advances with the anisotropic techniques described above it
seems feasible to develop fully self-consistent simulations of dendritic
solidification and related phenomena, and work on this task is underway.

\begin{acknowledgments}
This research has been supported by the EC under Contract No. HPMF-CT-2000-00800.
NRG was supported by an EPSRC CASE award. The computing facilities
were provided by the Centre for Scientific Computing of the University
of Warwick, with support from the JREI. 
\end{acknowledgments}

\bibliography{dlaref}

\begin{thebibliography}{31}
\expandafter\ifx\csname natexlab\endcsname\relax\def\natexlab#1{#1}\fi
\expandafter\ifx\csname bibnamefont\endcsname\relax
  \def\bibnamefont#1{#1}\fi
\expandafter\ifx\csname bibfnamefont\endcsname\relax
  \def\bibfnamefont#1{#1}\fi
\expandafter\ifx\csname citenamefont\endcsname\relax
  \def\citenamefont#1{#1}\fi
\expandafter\ifx\csname url\endcsname\relax
  \def\url#1{\texttt{#1}}\fi
\expandafter\ifx\csname urlprefix\endcsname\relax\def\urlprefix{URL }\fi
\providecommand{\bibinfo}[2]{#2}
\providecommand{\eprint}[2][]{\url{#2}}

\bibitem[{\citenamefont{Witten and Sander}(1981)}]{witten81}
\bibinfo{author}{\bibfnamefont{T.~A.} \bibnamefont{Witten}} \bibnamefont{and}
  \bibinfo{author}{\bibfnamefont{L.~M.} \bibnamefont{Sander}},
  \bibinfo{journal}{Phys.\ Rev.\ Lett.} \textbf{\bibinfo{volume}{47}},
  \bibinfo{pages}{1400} (\bibinfo{year}{1981}).

\bibitem[{\citenamefont{Plischke and R\'acz}(1984)}]{plischke84}
\bibinfo{author}{\bibfnamefont{M.}~\bibnamefont{Plischke}} \bibnamefont{and}
  \bibinfo{author}{\bibfnamefont{Z.}~\bibnamefont{R\'acz}},
  \bibinfo{journal}{Phys.\ Rev.\ Lett.} \textbf{\bibinfo{volume}{53}},
  \bibinfo{pages}{415} (\bibinfo{year}{1984}).

\bibitem[{\citenamefont{Coniglio and Zannetti}(1990)}]{coniglio90}
\bibinfo{author}{\bibfnamefont{A.}~\bibnamefont{Coniglio}} \bibnamefont{and}
  \bibinfo{author}{\bibfnamefont{M.}~\bibnamefont{Zannetti}},
  \bibinfo{journal}{Physica A} \textbf{\bibinfo{volume}{163}},
  \bibinfo{pages}{325} (\bibinfo{year}{1990}).

\bibitem[{\citenamefont{Somfai et~al.}(1999)\citenamefont{Somfai, Sander, and
  Ball}}]{somfai99}
\bibinfo{author}{\bibfnamefont{E.}~\bibnamefont{Somfai}},
  \bibinfo{author}{\bibfnamefont{L.~M.} \bibnamefont{Sander}},
  \bibnamefont{and} \bibinfo{author}{\bibfnamefont{R.~C.} \bibnamefont{Ball}},
  \bibinfo{journal}{Phys.\ Rev.\ Lett.} \textbf{\bibinfo{volume}{83}},
  \bibinfo{pages}{5523} (\bibinfo{year}{1999}).

\bibitem[{\citenamefont{Amitrano}(1989)}]{amitrano89}
\bibinfo{author}{\bibfnamefont{C.}~\bibnamefont{Amitrano}},
  \bibinfo{journal}{Phys.\ Rev.\ A} \textbf{\bibinfo{volume}{39}},
  \bibinfo{pages}{6618} (\bibinfo{year}{1989}).

\bibitem[{\citenamefont{Jensen et~al.}(2003)\citenamefont{Jensen, Mathiesen,
  and Procaccia}}]{jensen03}
\bibinfo{author}{\bibfnamefont{M.~H.} \bibnamefont{Jensen}},
  \bibinfo{author}{\bibfnamefont{J.}~\bibnamefont{Mathiesen}},
  \bibnamefont{and}
  \bibinfo{author}{\bibfnamefont{I.}~\bibnamefont{Procaccia}},
  \bibinfo{journal}{Phys.\ Rev.\ E} \textbf{\bibinfo{volume}{67}},
  \bibinfo{pages}{042402} (\bibinfo{year}{2003}).

\bibitem[{\citenamefont{Somfai et~al.}(2004)\citenamefont{Somfai, Goold, Ball,
  Devita, and Sander}}]{somfai04}
\bibinfo{author}{\bibfnamefont{E.}~\bibnamefont{Somfai}},
  \bibinfo{author}{\bibfnamefont{N.~R.} \bibnamefont{Goold}},
  \bibinfo{author}{\bibfnamefont{R.~C.} \bibnamefont{Ball}},
  \bibinfo{author}{\bibfnamefont{J.~P.} \bibnamefont{Devita}},
  \bibnamefont{and} \bibinfo{author}{\bibfnamefont{L.~M.}
  \bibnamefont{Sander}}, \bibinfo{journal}{Phys.\ Rev.\ E}
  \textbf{\bibinfo{volume}{70}}, \bibinfo{pages}{051403}
  (\bibinfo{year}{2004}).

\bibitem[{\citenamefont{Langer}(1980)}]{langer80}
\bibinfo{author}{\bibfnamefont{J.~S.} \bibnamefont{Langer}},
  \bibinfo{journal}{Rev.\ Mod.\ Phys.} \textbf{\bibinfo{volume}{52}},
  \bibinfo{pages}{1} (\bibinfo{year}{1980}).

\bibitem[{\citenamefont{Mullins and Sekerka}(1963)}]{mullins63}
\bibinfo{author}{\bibfnamefont{W.~W.} \bibnamefont{Mullins}} \bibnamefont{and}
  \bibinfo{author}{\bibfnamefont{R.~F.} \bibnamefont{Sekerka}},
  \bibinfo{journal}{J.\ Appl.\ Phys.} \textbf{\bibinfo{volume}{34}},
  \bibinfo{pages}{323} (\bibinfo{year}{1963}).

\bibitem[{\citenamefont{Nittmann et~al.}(1985)\citenamefont{Nittmann, Daccord,
  and Stanley}}]{nittmann85}
\bibinfo{author}{\bibfnamefont{J.}~\bibnamefont{Nittmann}},
  \bibinfo{author}{\bibfnamefont{G.}~\bibnamefont{Daccord}}, \bibnamefont{and}
  \bibinfo{author}{\bibfnamefont{H.~E.} \bibnamefont{Stanley}},
  \bibinfo{journal}{Nature} \textbf{\bibinfo{volume}{314}},
  \bibinfo{pages}{141} (\bibinfo{year}{1985}).

\bibitem[{\citenamefont{Matsushita et~al.}(1984)\citenamefont{Matsushita, Sano,
  Hayakawa, Honjo, and Sawada}}]{matsushita84}
\bibinfo{author}{\bibfnamefont{M.}~\bibnamefont{Matsushita}},
  \bibinfo{author}{\bibfnamefont{M.}~\bibnamefont{Sano}},
  \bibinfo{author}{\bibfnamefont{Y.}~\bibnamefont{Hayakawa}},
  \bibinfo{author}{\bibfnamefont{H.}~\bibnamefont{Honjo}}, \bibnamefont{and}
  \bibinfo{author}{\bibfnamefont{Y.}~\bibnamefont{Sawada}},
  \bibinfo{journal}{Phys.\ Rev.\ Lett.} \textbf{\bibinfo{volume}{53}},
  \bibinfo{pages}{286} (\bibinfo{year}{1984}).

\bibitem[{\citenamefont{Brady and Ball}(1984)}]{brady84}
\bibinfo{author}{\bibfnamefont{R.~M.} \bibnamefont{Brady}} \bibnamefont{and}
  \bibinfo{author}{\bibfnamefont{R.~C.} \bibnamefont{Ball}},
  \bibinfo{journal}{Nature} \textbf{\bibinfo{volume}{309}},
  \bibinfo{pages}{225} (\bibinfo{year}{1984}).

\bibitem[{\citenamefont{Niemeyer et~al.}(1984)\citenamefont{Niemeyer,
  Pietronero, and Wiesmann}}]{niemeyer84}
\bibinfo{author}{\bibfnamefont{L.}~\bibnamefont{Niemeyer}},
  \bibinfo{author}{\bibfnamefont{L.}~\bibnamefont{Pietronero}},
  \bibnamefont{and} \bibinfo{author}{\bibfnamefont{H.~J.}
  \bibnamefont{Wiesmann}}, \bibinfo{journal}{Phys.\ Rev.\ Lett.}
  \textbf{\bibinfo{volume}{52}}, \bibinfo{pages}{1033} (\bibinfo{year}{1984}).

\bibitem[{\citenamefont{Nittman and Stanley}(1987)}]{nittmann87}
\bibinfo{author}{\bibfnamefont{J.}~\bibnamefont{Nittman}} \bibnamefont{and}
  \bibinfo{author}{\bibfnamefont{H.~E.} \bibnamefont{Stanley}},
  \bibinfo{journal}{J.\ Phys.\ A:\ Math.\ Gen.} \textbf{\bibinfo{volume}{20}},
  \bibinfo{pages}{L1185} (\bibinfo{year}{1987}).

\bibitem[{\citenamefont{Ball and Somfai}(2002)}]{ball02prl}
\bibinfo{author}{\bibfnamefont{R.~C.} \bibnamefont{Ball}} \bibnamefont{and}
  \bibinfo{author}{\bibfnamefont{E.}~\bibnamefont{Somfai}},
  \bibinfo{journal}{Phys.\ Rev.\ Lett.} \textbf{\bibinfo{volume}{89}},
  \bibinfo{pages}{135503} (\bibinfo{year}{2002}).

\bibitem[{\citenamefont{Ball and Somfai}(2003)}]{ball03pre}
\bibinfo{author}{\bibfnamefont{R.~C.} \bibnamefont{Ball}} \bibnamefont{and}
  \bibinfo{author}{\bibfnamefont{E.}~\bibnamefont{Somfai}},
  \bibinfo{journal}{Phys.\ Rev.\ E} \textbf{\bibinfo{volume}{67}},
  \bibinfo{pages}{021401} (\bibinfo{year}{2003}).

\bibitem[{\citenamefont{Libbrecht and Tanusheva}(1999)}]{libbrecht99}
\bibinfo{author}{\bibfnamefont{K.~G.} \bibnamefont{Libbrecht}}
  \bibnamefont{and} \bibinfo{author}{\bibfnamefont{V.~M.}
  \bibnamefont{Tanusheva}}, \bibinfo{journal}{Phys.\ Rev.\ E}
  \textbf{\bibinfo{volume}{59}}, \bibinfo{pages}{3253} (\bibinfo{year}{1999}).

\bibitem[{\citenamefont{Huang and Glicksman}(1981)}]{huang81}
\bibinfo{author}{\bibfnamefont{S.~C.} \bibnamefont{Huang}} \bibnamefont{and}
  \bibinfo{author}{\bibfnamefont{M.~E.} \bibnamefont{Glicksman}},
  \bibinfo{journal}{Acta.\ Metall.} \textbf{\bibinfo{volume}{29}},
  \bibinfo{pages}{717} (\bibinfo{year}{1981}).

\bibitem[{\citenamefont{Corrigan et~al.}(1999)\citenamefont{Corrigan, Koss,
  LaCombe, de~Jager, Tennenhouse, and Glicksman}}]{corrigan99}
\bibinfo{author}{\bibfnamefont{D.~P.} \bibnamefont{Corrigan}},
  \bibinfo{author}{\bibfnamefont{M.~B.} \bibnamefont{Koss}},
  \bibinfo{author}{\bibfnamefont{J.~C.} \bibnamefont{LaCombe}},
  \bibinfo{author}{\bibfnamefont{K.~D.} \bibnamefont{de~Jager}},
  \bibinfo{author}{\bibfnamefont{L.~A.} \bibnamefont{Tennenhouse}},
  \bibnamefont{and} \bibinfo{author}{\bibfnamefont{M.~E.}
  \bibnamefont{Glicksman}}, \bibinfo{journal}{Phys.\ Rev.\ E}
  \textbf{\bibinfo{volume}{60}}, \bibinfo{pages}{7217} (\bibinfo{year}{1999}).

\bibitem[{\citenamefont{Russel et~al.}(1997)\citenamefont{Russel, Chaikin, Zhu,
  Meyer, and Rogers}}]{russel97}
\bibinfo{author}{\bibfnamefont{W.~B.} \bibnamefont{Russel}},
  \bibinfo{author}{\bibfnamefont{P.~M.} \bibnamefont{Chaikin}},
  \bibinfo{author}{\bibfnamefont{J.}~\bibnamefont{Zhu}},
  \bibinfo{author}{\bibfnamefont{M.~V.} \bibnamefont{Meyer}}, \bibnamefont{and}
  \bibinfo{author}{\bibfnamefont{R.}~\bibnamefont{Rogers}},
  \bibinfo{journal}{Langmuir} \textbf{\bibinfo{volume}{13}},
  \bibinfo{pages}{3871} (\bibinfo{year}{1997}).

\bibitem[{\citenamefont{Zhu et~al.}(1997)\citenamefont{Zhu, Li, Rogers, Meyer,
  Ottewill, Russel, and Chaikin}}]{zhu97}
\bibinfo{author}{\bibfnamefont{J.}~\bibnamefont{Zhu}},
  \bibinfo{author}{\bibfnamefont{M.}~\bibnamefont{Li}},
  \bibinfo{author}{\bibfnamefont{R.}~\bibnamefont{Rogers}},
  \bibinfo{author}{\bibfnamefont{W.}~\bibnamefont{Meyer}},
  \bibinfo{author}{\bibfnamefont{R.~H.} \bibnamefont{Ottewill}},
  \bibinfo{author}{\bibfnamefont{W.~B.} \bibnamefont{Russel}},
  \bibnamefont{and} \bibinfo{author}{\bibfnamefont{P.~M.}
  \bibnamefont{Chaikin}}, \bibinfo{journal}{Nature}
  \textbf{\bibinfo{volume}{387}}, \bibinfo{pages}{883} (\bibinfo{year}{1997}).

\bibitem[{\citenamefont{Glicksman and Marsh}(1993)}]{glicksman93}
\bibinfo{author}{\bibfnamefont{M.~E.} \bibnamefont{Glicksman}}
  \bibnamefont{and} \bibinfo{author}{\bibfnamefont{S.~P.} \bibnamefont{Marsh}},
  in \emph{\bibinfo{booktitle}{Handbook of Crystal Growth}}, edited by
  \bibinfo{editor}{\bibfnamefont{D.~T.~J.} \bibnamefont{Hurle}}
  (\bibinfo{publisher}{Elsevier}, \bibinfo{address}{Amsterdam},
  \bibinfo{year}{1993}), p. \bibinfo{pages}{1077}.

\bibitem[{\citenamefont{Kobayashi}(1993)}]{kobayashi93}
\bibinfo{author}{\bibfnamefont{R.}~\bibnamefont{Kobayashi}},
  \bibinfo{journal}{Physica D} \textbf{\bibinfo{volume}{63}},
  \bibinfo{pages}{410} (\bibinfo{year}{1993}).

\bibitem[{\citenamefont{George and Warren}(2002)}]{george02}
\bibinfo{author}{\bibfnamefont{W.~L.} \bibnamefont{George}} \bibnamefont{and}
  \bibinfo{author}{\bibfnamefont{J.~A.} \bibnamefont{Warren}},
  \bibinfo{journal}{J.\ Comp.\ Phys.} \textbf{\bibinfo{volume}{177}},
  \bibinfo{pages}{264} (\bibinfo{year}{2002}).

\bibitem[{\citenamefont{Meakin}(1987)}]{meakin87}
\bibinfo{author}{\bibfnamefont{P.}~\bibnamefont{Meakin}},
  \bibinfo{journal}{Phys.\ Rev.\ A} \textbf{\bibinfo{volume}{36}},
  \bibinfo{pages}{332} (\bibinfo{year}{1987}).

\bibitem[{\citenamefont{Ball et~al.}(1985)\citenamefont{Ball, Brady, Rossi, and
  Thompson}}]{ball85prl}
\bibinfo{author}{\bibfnamefont{R.~C.} \bibnamefont{Ball}},
  \bibinfo{author}{\bibfnamefont{R.~M.} \bibnamefont{Brady}},
  \bibinfo{author}{\bibfnamefont{G.}~\bibnamefont{Rossi}}, \bibnamefont{and}
  \bibinfo{author}{\bibfnamefont{B.~R.} \bibnamefont{Thompson}},
  \bibinfo{journal}{Phys.\ Rev.\ Lett.} \textbf{\bibinfo{volume}{55}},
  \bibinfo{pages}{1406} (\bibinfo{year}{1985}).

\bibitem[{\citenamefont{Ball and Brady}(1985)}]{ball85jphysa}
\bibinfo{author}{\bibfnamefont{R.~C.} \bibnamefont{Ball}} \bibnamefont{and}
  \bibinfo{author}{\bibfnamefont{R.~M.} \bibnamefont{Brady}},
  \bibinfo{journal}{J.\ Phys.\ A} \textbf{\bibinfo{volume}{18}},
  \bibinfo{pages}{L809} (\bibinfo{year}{1985}).

\bibitem[{\citenamefont{Tolman and Meakin}(1989)}]{tolman89}
\bibinfo{author}{\bibfnamefont{S.}~\bibnamefont{Tolman}} \bibnamefont{and}
  \bibinfo{author}{\bibfnamefont{P.}~\bibnamefont{Meakin}},
  \bibinfo{journal}{Phys.\ Rev.\ A} \textbf{\bibinfo{volume}{40}},
  \bibinfo{pages}{428} (\bibinfo{year}{1989}).

\bibitem[{\citenamefont{Bowler and Ball}(2004)}]{bowler04}
\bibinfo{author}{\bibfnamefont{N.~E.} \bibnamefont{Bowler}} \bibnamefont{and}
  \bibinfo{author}{\bibfnamefont{R.~C.} \bibnamefont{Ball}}
  (\bibinfo{year}{2004}), \eprint{cond-mat/0404650}.

\bibitem[{\citenamefont{Ball et~al.}(2002)\citenamefont{Ball, Bowler, Sander,
  and Somfai}}]{ball02pre}
\bibinfo{author}{\bibfnamefont{R.~C.} \bibnamefont{Ball}},
  \bibinfo{author}{\bibfnamefont{N.~E.} \bibnamefont{Bowler}},
  \bibinfo{author}{\bibfnamefont{L.~M.} \bibnamefont{Sander}},
  \bibnamefont{and} \bibinfo{author}{\bibfnamefont{E.}~\bibnamefont{Somfai}},
  \bibinfo{journal}{Phys.\ Rev.\ E} \textbf{\bibinfo{volume}{66}},
  \bibinfo{pages}{026109} (\bibinfo{year}{2002}).

\bibitem[{\citenamefont{Barker and Ball}(1990)}]{barker90}
\bibinfo{author}{\bibfnamefont{P.~W.} \bibnamefont{Barker}} \bibnamefont{and}
  \bibinfo{author}{\bibfnamefont{R.~C.} \bibnamefont{Ball}},
  \bibinfo{journal}{Phys.\ Rev.\ A} \textbf{\bibinfo{volume}{42}},
  \bibinfo{pages}{R6289} (\bibinfo{year}{1990}).

\end{thebibliography}

\end{document}